\documentclass[11pt]{article}
\usepackage{epsfig}
\newcommand{\be}{\begin{equation}}
\newcommand{\ee}{\end{equation}}
\newcommand{\ba}{\begin{eqnarray}}
\newcommand{\ea}{\end{eqnarray}}
\def\L{{\cal L}}
\newcommand{\e}{{\rm e}}
\newcommand{\ep}{\epsilon}
\newcommand{\p}{\partial}
\newcommand{\vev}[1]{\left\langle #1 \right\rangle}

\newcommand{\dslash}{\hbox{$\partial$\kern-0.5em\raise0.3ex\hbox{/}}}
\def\slash#1{\hbox{$#1$\kern-0.5em\raise0.3ex\hbox{/}}}
\newcommand{\A}{{\cal A}}
\newcommand{\cbar}{\bar{c}}
\begin{document}
\begin{titlepage}
\rightline{KOBE-TH-01-05}
\rightline{\tt hep-th/0108217}
\vspace{.5cm}
\begin{center}
{\LARGE Off-shell renormalization of \\
\vspace{0.3cm} the abelian Higgs model in the unitary gauge}\\
 \renewcommand{\thefootnote}{\fnsymbol{footnote}}
\vspace{1cm} Hidenori SONODA\footnote[2]{E-mail: {\tt
sonoda@phys.sci.kobe-u.ac.jp}}\\
\renewcommand{\thefootnote}{\arabic{footnote}}
\vspace{.2cm}
Physics Department, Kobe University, Kobe 657-8501, Japan\\
\vspace{.2cm} 
August 2001\\
\vspace{.2cm}
PACS numbers: 11.10.Gh, 11.15.-q\\
Keywords: renormalization, gauge field theories
\end{center}
\vspace{.3cm}
\begin{abstract}
We discuss the off-shell renormalization properties of the abelian
Higgs model in the unitary gauge.  The model is not renormalizable
according to the usual power counting rules.  In this paper, however,
we show that with a proper choice of interpolating fields for the
massive photon and the Higgs particle, their off-shell Green functions
can be renormalized.  An analysis of the nature of the extra
singularities in the unitary gauge is given, and a recipe for the
off-shell renormalization is provided.
\end{abstract}
\end{titlepage}

\section{Introduction}

The choice of the unitary gauge for Higgs theories is usually
dismissed on account of the non-renormalizability due to power
counting.  This is not a happy situation since, first of all, the
unitary gauge is physically a nice gauge to work with; it is devoid of
gauge degrees of freedom, and only the physical degrees of freedom are
retained.  It is difficult to dismiss this gauge.  We also believe
that the unitary gauge should make sense physically, for it is
formally equivalent to any other manifestly renormalizable gauges.
What is most troubling is that we have no clear idea of what is
causing the extra divergences which are unique to the unitary gauge.
It has been shown that the unitary gauge can be used for the loop
calculation of physical quantities such as the S-matrix elements
\cite{Dublin}.  Nevertheless, the situation is still not satisfying,
because the cancellations of the extra UV divergences occur
miraculously as we take the mass shell limit of the Green functions.
What is missing is a good understanding of the nature of the extra UV
divergences in the unitary gauge.

The purpose of this paper is to provide a good understanding of the
renormalization properties of the unitary gauge.  Our conclusion is
that the unitary gauge is as renormalizable as the renormalizable
gauges.

There are two causes for the extra divergences in the unitary gauge.
The first is an improper choice of the interpolating fields of the
elementary particles.  The second is the compositeness of the
interpolating fields.  The first implies that the simplest choice of
the fields for the massive gauge boson and Higgs particle are not
``good'' local fields.  Here, a good local field has a well defined
scale dimension, and it mixes with a finite number of other local
fields under the renormalization group transformations.  The second
implies that the off-shell Green functions are UV finite only in the
coordinate space for all distinct points.  The proper interpolating
fields of the massive photon and Higgs particle turn out to be
composite fields with scale dimensions $2$ or higher, and when two or
more of them coincide in space, a non-integrable singularity is
generated.  The Fourier transform of the Green function thus acquires
additional UV singularities which cannot be removed by multiplicative
renormalization of the fields.  These singularities do not affect the
S-matrix, since only the asymptotic limit of infinite spatial
separation of the interpolating fields is relevant for the S-matrix.
The same singularities would appear even in the renormalizable gauges
if we chose higher dimensional composite fields as the interpolating
fields.

This paper is an extension of the previous paper \cite{S} in which the
off-shell renormalizability of the massive QED was explained.  The
abelian Higgs model, discussed in this paper, is more non-trivial due
to the presence of the Higgs particle.  A further extension to the
non-abelian Higgs theories is left for future.  The organization of
the present paper is as follows.  In sect.~2, we review the relation
of the unitary gauge to the other renormalizable gauges.  In sect.~3,
concrete examples of the 1-loop UV divergences in the unitary gauge
are given, and possible remedies for removing them are briefly
indicated.  In sect.~4, following the observations made in sect.~3, we
discuss the proper choice of interpolating fields.  In sect.~5, for
completeness of the paper, we remind the reader of the renormalization
conditions suitable for the unitary gauge.  Then, in sect.~6, we
analyze the structure of the UV divergences of the Green functions.
Sects.~3, 4, and 6 constitute the main part of this paper.  We briefly
summarize how to calculate the S-matrix elements in the unitary gauge
in sect.~7 before we conclude the paper in sect.~8.  Two appendices
are given to give more details of the derivations in the main text.

Throughout the paper we use the euclidean metric and dimensional
regularization for the $D \equiv 4 - \ep$ dimensional space.

\section{Review --- equivalence of the unitary gauge with the
renormalizable gauges}

As was first emphasized in ref.~\cite{DJ} and more recently in
ref.~\cite{TVK}, the unitary gauge is not strictly speaking a choice
of gauge.  It is instead a choice of field variables (polar variables
in the language of ref.~\cite{TVK}) which separate the gauge
independent degrees of freedom from the gauge dependent ones.  In this
section we review the relation of the unitary gauge to the
renormalizable $R_\xi$ gauge \cite{FLS}.  Our presentation is based
upon the old analyses made in refs.~\cite{LZ, DJ}.  Our summary is
somewhat simpler thanks to the use of dimensional regularization.

The gauge invariant quantities can be calculated in any gauge, either
in the unitary gauge or in any renormalizable gauge such as the
$R_\xi$ gauge.  It is the purpose of this section to give a formal
proof of this statement.  (See also \cite{TVK} for a more perturbative
derivation.)

The lagrangian of the abelian Higgs model is given by
\ba \L_\xi &=& {1 \over 4} F^2 + \left(\p_\mu \phi^* + i e A_\mu
\phi^*\right)\left( \p_\mu \phi - i e A_\mu \phi \right)
+ {\lambda \over 4} \left(|\phi|^2 - {v^2 \over 2}\right)^2 \nonumber\\
&& + {1 \over 2 \xi} \left( \p \cdot A - \xi e v \chi \right)^2 +
\p_\mu \cbar \p_\mu c + \xi e^2 v \rho' \cbar c \label{Rxi} \ea
where the real fields $\rho'$, $\chi$ are defined by $\phi \equiv {1
\over \sqrt{2}} (\rho' + i \chi)$, and $c,\cbar$ are the FP ghosts.
This lagrangian is invariant under the BRST transformation defined by
\ba && \delta_\eta A_\mu = \eta \p_\mu c \nonumber\\ &&\delta_\eta \phi = i
e \eta c \phi, \quad \delta_\eta \phi^* = - i e \eta c \phi^*
\nonumber\\ && \delta_\eta c = 0, \quad \delta_\eta \cbar = {1 \over \xi}
\eta \left( \p \cdot A - \xi e v \chi \right) \ea
where $\eta$ is an arbitrary Grassmann number.  

To go to the unitary gauge, we make the following change of variables:
\be B_\mu \equiv A_\mu - \p_\mu \varphi,\quad {1 \over \sqrt{2}} ~\rho
\e^{i \varphi} \equiv \phi
\label{polar}\ee
What is nice about the dimensional regularization is that the jacobian
of the change of variables is unity.  Hence, the lagrangian for the
new fields is obtained simply by substituting (\ref{polar}) into the
lagrangian (\ref{Rxi}):
\be \L_\xi = \L_U + \Delta \L \ee
where $\L_U$ is the lagrangian for the unitary gauge
\be \L_U = {1 \over 4} \left( \p_\mu B_\nu - \p_\nu B_\mu \right)^2 +
{1 \over 2} \left( \left(\p_\mu \rho \right)^2 + e^2 B_\mu^2 \rho^2
\right) + {\lambda \over 16} \left( \rho^2 - v^2 \right)^2 \ee
and
\be \Delta \L = {1 \over 2 \xi e^2} \left( e \p \cdot B + \p^2
\varphi - \xi e^2 v \rho \sin \varphi \right)^2 + \p_\mu \cbar \p_\mu
c + \xi e^2 v \rho \cos \varphi \cbar c \ee
Since $\L_U$ is gauge invariant (hence BRST invariant), the difference
$\Delta \L = \L_\xi - \L_U$ must be also BRST invariant.  Since the
gauge invariant quantities do not depend on $\varphi$, $c$, and
$\cbar$, we can integrate them out first.  We now show that
\be Z \equiv \int [d\varphi dc d\cbar] \e^{- \int \Delta \L} \ee
is independent of $\rho, B_\mu$.  To show this we compute the
functional derivatives ${\delta Z \over \delta \rho (x)}$, ${\delta Z
\over \delta B_\mu (x)}$.  First, we find
\be \eta {\delta Z \over \delta \rho (x)} = \int [d\varphi dc d\cbar]
\xi e v \delta_\eta \left( \cbar \sin \varphi \right) \cdot \e^{- \int
\Delta \L} = 0 \ee
since $\delta_\eta \delta L = 0$, and the measure of integration is
BRST invariant.  Similarly, we find
\be \eta {\delta Z \over \delta B_\mu (x)} = \int [d\varphi dc d\cbar]
\p_\mu \delta_\eta \cbar \cdot \e^{- \int \Delta \L} = 0 \ee

Thus, $Z$ is independent of $\rho$, $B_\mu$, and we obtain the
desired equality
\be \int [d B_\mu d \rho d\varphi dc d\cbar] ~B_\mu ... \rho ... ~\e^{-
\int \L_\xi} = \int [d B_\mu d\rho ] ~B_\mu ... \rho ... ~\e^{- \int
\L_U} \ee
For those gauge invariant quantities that depend only on the gauge
invariant fields $B_\mu, \rho$, we can use either the $R_\xi$ gauge or
the unitary gauge.  We get the same result.  Of course, this must be
true if we recall how the lagrangian of the $R_\xi$ gauge can be
obtained using the method of Faddeev and Popov \cite{FP}.
Alternatively the equivalence can be shown by taking the formal limit
of $\xi \to \infty$.  The advantage of the above proof is that it is
consistent with regularization, and hence the proof can incorporate
renormalization.

\section{Examples of the UV divergences in the unitary gauge}

To make our discussions concrete, we show four examples of 1-loop
calculations in the unitary gauge.  We renormalize only the parameters
of the model and leave all the fields unrenormalized:
\be e_0^2 = {e^2 \over Z_3}, \quad (\lambda v^2)_0 = Z_m \lambda
v^2,\quad \lambda_0 = Z_\lambda \lambda + z \label{parameters}\ee
where the parameters with suffix $0$ are bare parameters, and $z$
depends only on $e^2$ but not on $\lambda$.\footnote{Note that
$(\lambda v^2)_0 \ne \lambda_0 v_0^2$.  The shift $v_0$ can be chosen
arbitrarily.}  Shifting $\rho$ by an arbitrary constant $v_0 = Z_v v$,
i.e., $\rho \to \rho + v_0$, we obtain the following lagrangian
\ba \L_U &=& {1 \over 4} (\p_\mu B_\nu - \p_\nu B_\mu)^2  +
{1 \over 2} \left( (\p_\mu \rho)^2 + {e^2 \over Z_3} B_\mu^2 (Z_v v +
\rho)^2 \right)\nonumber\\
&&  - {1 \over 8} Z_m \lambda v^2 (Z_v v + \rho)^2 + {1
\over 16} (Z_\lambda \lambda + z) (Z_v v + \rho)^4 \ea
The renormalization of the parameters is gauge invariant, and we can
choose the renormalization constants the same as in any renormalizable
gauge:
\ba Z_3 &=& 1 + {1 \over (4\pi)^2} {2 \over \ep} \left( - {e^2 \over
3}\right) \label{z3}\\ Z_m &=& 1 + {1 \over (4\pi)^2} {2 \over \ep}
(\lambda - 3 e^2)\label{zm}\\ Z_\lambda &=& 1 + {1 \over (4\pi)^2} {2
\over \ep} \left( {5 \over 2} \lambda - 6 e^2\right), \quad z = {1
\over (4\pi)^2} {2 \over \ep} (12 e^4) \label{zlambda}\ea
The constant $Z_v$ is chosen for convenience so that the tadpole
vanishes, $\vev{\rho} = 0$:
\be Z_v = 1 + {1 \over (4\pi)^2} {2 \over \ep} \left( {3 e^2 \over
2}\right) + (\rm{finite}) \ee

We consider four Green functions: the photon propagator $\vev{B_\mu
B_\nu}$, the Higgs propagator $\vev{\rho \rho}$, and then two
three-point functions $\vev{\rho B_\mu B_\nu}, \vev{\rho \rho
\rho}$.\footnote{Note $\vev{\rho~\rho~B_\mu}=0$ and $\vev{B_\alpha
B_\beta B_\gamma} = 0$ due to the charge conjugation invariance.}  The
following UV divergences are found at 1-loop:
\ba &&UV \vev{B_\mu (k) B_\nu} = \ln \Lambda \left[ - {e^2 \over 3}
\vev{B_\mu (k) B_\nu} + {1 \over e^2 v^4} k_\mu k_\nu \right] \\ &&UV
\vev{\rho (p) \rho} = \ln \Lambda \left[ \left(3 e^2 - {\lambda \over
2}\right) \vev{\rho (p) \rho} + {1 \over 2 v^2} \right]\\ && UV
\vev{\rho ~B_\mu(k_1) B_\nu(k_2)} = \ln \Lambda \Bigg[ \left( - {e^2
\over 3} + {1 \over 2} \left(3 e^2 - {\lambda \over 2}\right)\right)
\vev{\rho ~B_\mu ~B_\nu} \nonumber\\ && \qquad\qquad\qquad\qquad\qquad
+ {4 \over e^2 v^5 } k_{1\mu} k_{2\nu} \vev{\rho~ \rho(k_1+k_2)}
\Bigg] \\&& UV \vev{\rho (k_1)~ \rho (k_2)~ \rho} = \ln \Lambda \Bigg[
{3 \over 2} \left(3 e^2 - {\lambda \over 2}\right) \vev{\rho ~\rho
~\rho} \nonumber\\ && \qquad + {\lambda \over 4} \left(\vev{{\rho^2
\over 2v} ~\rho ~\rho} + \vev{\rho~ {\rho^2 \over 2v}~ \rho} +
\vev{\rho~ \rho~ {\rho^2 \over 2v}}\right) \nonumber\\ &&\qquad - {1
\over v^3} \left(\vev{\rho(k_1) ~\rho} + \vev{\rho(k_2) ~\rho} +
\vev{\rho (k_1+k_2) ~\rho}\right)\Bigg] \ea
where $UV$ denotes that we take only the UV singular part, and $\ln
\Lambda \equiv {1 \over (4\pi)^2} {2 \over \ep}$.

Clearly, multiplicative renormalization of the fields removes the
first term of singularity for each Green function.  But after the wave
function renormalization we are still left with the following UV
divergences:
\ba &&UV \vev{B_\mu (k) B_\nu} = \ln \Lambda~ {1 \over e^2 v^4} k_\mu
k_\nu \label{BB}\\ &&UV \vev{\rho (p) \rho} = \ln \Lambda ~{1 \over 2
v^2} \label{rhorho}\\ &&UV \vev{\rho ~B_\mu(k_1) B_\nu(k_2)} = \ln
\Lambda ~{4 \over e^2 v^5} k_{1\mu} k_{2\nu} \vev{\rho ~\rho
(k_1+k_2)}
\label{rhoBB}\\ && UV \vev{\rho (k_1) ~\rho (k_2) ~\rho} = \ln \Lambda
\Bigg[ {\lambda \over 4} \left(\vev{{\rho^2 \over 2v} ~\rho ~\rho} +
\vev{\rho ~{\rho^2 \over 2v} ~\rho} + \vev{\rho ~\rho ~{\rho^2 \over
2v}}\right)\nonumber\\ && \quad - {1 \over v^3} \left(\vev{\rho(k_1)
~\rho} + \vev{\rho(k_2) ~\rho} + \vev{\rho (k_1+k_2)
~\rho}\right)\Bigg]
\label{rhorhorho}\ea
The local divergences in Eqs.~(\ref{BB},\ref{rhorho}) imply that the
scale dimensions of $B_\mu$, $\rho$ are actually $3$ and $2$,
respectively, to the contrary to the naively expected scale dimension
1 for both fields.\footnote{The tree-level propagator of $B_\mu$
already implies $2$ for its scale dimension.  This is the very cause
of non-renormalizability by power counting.}  The non-local divergence
given by Eq.~(\ref{rhoBB}) is at first sight difficult to remove.
From the scale dimension $3$ of $B_\mu$ found above, we expect the
following operator product expansion (OPE) in coordinate space
\be B_\mu (x) B_\nu (0) \sim {1 \over x^4} \left( \delta_{\mu\nu} - 4
{x_\mu x_\nu \over x^2} \right) {1 \over e^2 v^3} \rho (0) \ee
but this gives a singularity of the form
\be \ln \Lambda~\delta_{\mu\nu} {1 \over e^2 v^3} \vev{\rho ~\rho
(k_1+k_2)} \ee
which is different from the singularity of Eq.~(\ref{rhoBB}).  The
same singularity is actually contained in the Green function with
$B_\mu$ replaced by $B_\mu {\rho \over v}$:
\ba && UV \left[\vev{\rho \left(B_\mu {\rho \over v}\right)(k_1)
B_\nu(k_2)} + \vev{\rho~ B_\mu (k_1) \left(B_\nu {\rho \over
v}\right)(k_2)}\right] \nonumber\\ &=& \ln \Lambda \Bigg[ - {20 e^2
\over 3} \vev{\rho ~B_\mu ~B_\nu} - {11 e^2 \over 3} \left( \vev{\rho
\left(B_\mu {\rho \over v}\right) B_\nu} + \vev{\rho~B_\mu \left(B_\nu
{\rho \over v}\right)} \right) \nonumber\\ && \qquad - {8 \over e^2
v^5} k_{1\mu} k_{2\nu} \vev{\rho~\rho(k_1+k_2)} \Bigg]\ea
This suggests that we must consider a linear combination of $B_\mu$
and $B_\mu {\rho \over v}$ for renormalization.  

Out of the two singularities given by Eq.~(\ref{rhorhorho}), the first
implies that $\rho$ mixes with ${1 \over v}{\rho^2 \over 2}$ under
renormalization.  The second implies also the necessity of considering
a linear combination of $\rho$ and ${\rho^2 \over 2v}$.  This is
because the Green function of ${\rho^2 \over 2v}$ has the singularity
\ba && UV \left[\vev{{\rho^2 \over 2v}~\rho~\rho} + \vev{\rho~{\rho^2
\over 2v}~\rho} + \vev{\rho~\rho~{\rho^2 \over 2v}}\right] \nonumber\\
&&= \ln \Lambda \Bigg[ - {9 \lambda \over 4} \vev{\rho~\rho~\rho} - {3
\lambda \over 4} \left(\vev{{\rho^2 \over 2v} ~\rho ~\rho} + \vev{\rho
~{\rho^2 \over 2v} ~\rho} + \vev{\rho ~\rho ~{\rho^2 \over
2v}}\right)\nonumber\\ && + {1 \over v^3} \left(\vev{\rho(k_1) ~\rho} +
\vev{\rho(k_2) ~\rho} + \vev{\rho (k_1+k_2) ~\rho}\right)\Bigg] \ea
and the last singularity can cancel the second singularity of
Eq.~(\ref{rhorhorho}).

We are not quite done yet with renormalizing the fields $B_\mu$ and
$\rho$, but we have obtained enough clues to the right way of doing
renormalization.  To summarize the preliminary consideration in this
section, we have found that the scale dimensions of $B_\mu$ and $\rho$
are $3$ and $2$, respectively, and that for renormalization we are
compelled to consider a linear combination of $B_\mu$ and $B_\mu {\rho
\over v}$ for the massive photon and that of $\rho$ and ${\rho^2 \over
2v}$ for the Higgs.

\section{Choice of interpolating fields}

Naively, both $B_\mu$ and $\rho$ are elementary fields of scale
dimension 1.  But our observations in the previous section indicate to
the contrary: the correct scale dimensions of $B_\mu$ and $\rho$ are 3
and 2, respectively.  Furthermore, the fields $B_\mu$ and $\rho$
cannot be renormalized multiplicatively, and we must consider a linear
combination of $B_\mu$ and $B_\mu {\rho \over v}$ for the massive
photon and that of $\rho$ and ${\rho^2 \over 2v}$ for the Higgs.  To
understand the necessity of considering such linear combinations, let
us consider $\rho$ further.  In terms of the original field variables
$\phi$ and $\phi^*$, $\rho$ is given as
\be \rho = - v_0 + \sqrt{ 2 \phi^* \phi }  \ee
Taking $\phi^*\phi - {v_0^2 \over 2}$ as a small field, we can expand
\be \rho = v_0 \left( {1 \over 2} \left({2 \phi^* \phi \over
v_0^2}-1\right) - {1 \over 8} \left({2 \phi^* \phi \over
v_0^2}-1\right)^2 + ... \right) \ee
Hence, in terms of $\phi^*\phi$, $\rho$ is obtained as an infinite
series.  The field $\phi^*\phi$ translates to a well defined local
field of scale dimension $2$ upon quantization, but $\rho$ does not.
$\rho$ is a linear combination of an infinite number of local fields
$\phi^* \phi$, and thus it will not have a well defined scale
dimension.  It is simply impossible to quantize $\rho$.

In the unitary gauge, one might naively expect that any local
polynomials of $\rho$ and $B_\mu$ are ``good'' local fields with
definite scale dimensions.  But we have found this is not the case.
``Good'' fields are local polynomials in terms of the original field
variables $A_\mu$, $\phi$, and $\phi^*$.  Neither $\rho$ nor $B_\mu$
meets this criterion.  To avoid inconvenient mixing with lower
dimensional fields, we must choose interpolating fields of the
smallest scale dimensions.  Hence, the interpolating field of the
Higgs particle must be a gauge invariant scalar of the smallest
dimension which is invariant under charge conjugation.  Similarly, the
interpolating field of the massive photon must be a gauge invariant
vector of the smallest dimension that flips sign under charge
conjugation.  Such fields are unique, and are readily given by $\phi^*
\phi$ for the Higgs and the conserved charge current $J_\mu \equiv
\phi^* i (D_\mu - \overleftarrow{D}_\mu)\phi$ for the massive photon,
where $D_\mu$ is the covariant derivative.  These interpolating fields
have been considered to explain the absence of any qualitative
difference between the confining phase and Higgs phase of the standard
model \cite{AF}.\footnote{The existence of these gauge invariant
interpolating fields is essential for understanding the physical
meaning of the Coleman-Weinberg effect \cite{CW} or equivalently the
first-order nature of the BCS transition \cite{H}.}

To summarize the above discussion, we have found that the
interpolating field of the massive photon is given by $\A_\mu$, where
\be e_0 v_0^2 \A_\mu \equiv J_\mu \equiv \phi^* i (D_\mu -
\overleftarrow{D}_\mu)\phi = e_0 v_0^2 \cdot B_\mu \left( 1 + {\rho
\over v_0} \right)^2 \ee
The interpolating field of the Higgs is given by $\Phi$, where
\be v_0 \Phi \equiv \phi^* \phi - {v_0^2 \over 2} = v_0 \cdot \rho
\left( 1 + {\rho \over 2 v_0} \right) \ee

Already at this point it is obvious why the off-shell Green functions
of $\A_\mu$ and $\Phi$ should be renormalizable.  According to the
equivalence of the unitary gauge to any renormalizable gauge (reviewed
in sect.~2), the Green functions of $B_\mu$ and $\rho$ are the same in
the unitary and renormalizable gauges.  In the renormalizable gauges,
$\A_\mu$ and $\Phi$ are renormalizable
multiplicatively.\footnote{Strictly speaking $\Phi$ mixes with the
identity field.  In the renormalizable gauges $\A_\mu$ and $\Phi$ will
also mix with BRST trivial composite fields.}  Hence, also in the
unitary gauge, $\A_\mu$ and $\Phi$ are renormalizable.

Whenever fields of dimension 2 or higher are involved, the UV
singularities of the Green functions are not completely UV finite for
a good reason.  All the Green functions are UV finite in coordinate
space for distinct space points, but when two or more fields coincide
in space, we find unintegrable singularities which give rise to UV
divergences in the Fourier transform of the Green functions.  The
situation is the same in any renormalizable gauge.  In sect.~6 we
determine the structure of the expected UV divergences.

\section{Renormalization conditions}

This section is meant only for completeness.  The renormalization
constants for the parameters are introduced as in
Eqs.~(\ref{parameters}):
\be e_0^2 = {e^2 \over Z_3}, \quad
\left( \lambda v^2 \right)_0 = Z_m \lambda v^2, \quad
\lambda_0 = Z_\lambda \lambda + z \ee
where $e^2$, $\lambda v^2$, and $\lambda$ are renormalized parameters.

There are many ways of renormalizing the theory.  Let us describe only
two examples here.  First, in the MS scheme the renormalization
constants $Z_3-1$, $Z_m-1$, $Z_\lambda -1$, and $z$ have only the pole
parts in $\ep$.  Second, in the physical renormalization scheme, the
renormalization constants are defined by the following three
conditions:
\begin{itemize}
\item $m^2 \equiv e^2 v^2$ is the physical photon mass
squared. Namely, the propagator $\vev{\A_\mu (k) \A_\nu}$ has a pole
at $k^2 = - m^2$.
\item $m_H^2 \equiv {1 \over 2} \lambda v^2$ is the physical Higgs
mass squared.  Namely, the propagator $\vev{\Phi (p) \Phi}$ has a pole
at $p^2 = - m_H^2$.  (If $m_H^2 > 4 m^2$ and the Higgs is unstable, we
choose $m_H^2$ as the center of a smeared pole.)
\item The forward scattering amplitude of two Higgs bosons at zero
kinetic energy is $6 \lambda$.
\end{itemize}
There are many substitutes for the third condition.  We may
alternatively specify a value of the forward scattering amplitude of
two photons at zero energy.

\section{General analysis of the structure of the Green functions}

In this section we first study the multiplicative renormalization
properties of the interpolating fields $\A_\mu$ and $\Phi$.  Then, we
analyze the transverse nature of the Green functions involving
$\A_\mu$.  Finally, using scaling arguments, we determine the UV
divergences of the Green functions of $\A_\mu$ and $\Phi$ that are
left after multiplicative renormalization.

\subsection{Renormalization of the interpolating fields}

We first consider the renormalized conserved current $J_{r,\mu}$ and
the renormalized composite field $(\phi^* \phi)_r$.  Once they are
given, the renormalized $\A_{r,\mu}$ and $\Phi_r$ are defined by
\be e v^2 \A_{r,\mu} = J_{r,\mu}, \quad v \Phi_r = (\phi^* \phi)_r -
{v^2 \over 2} \ee

The renormalization of the conserved current $J_\mu$ is well known.
In terms of the bare fields the equation of motion is given by
\be \p_\nu F_{\nu\mu} = e_0 J_\mu \ee
where $F_{\nu\mu} \equiv \p_\nu B_\mu - \p_\mu B_\nu$ is the usual
field strength.  Since the field strength is renormalized by ${1 \over
\sqrt{Z_3}} F_{\nu\mu}$, the current is renormalized by
\be J_{r,\mu} \equiv {1 \over Z_3} J_\mu \ee

To renormalize $\phi^* \phi$, we recall it is conjugate to the mass
parameter $(\lambda v^2)_0$, i.e., it is given by the derivative of
the lagrangian with respect to the parameter: 
\be \phi^* \phi = - 4 {\p \L_U \over \p (\lambda v^2)_0} \ee
Therefore, the renormalized $\phi^* \phi$ is obtained by replacing the
bare parameter $(\lambda v^2)_0$ by the renormalized $\lambda v^2$.
Hence, the renormalization is done by
\be (\phi^* \phi)_r = Z_m \phi^* \phi \ee
Actually this is not quite correct, since $\phi^* \phi$ mixes with the
identity field under renormalization.  The correct prescription is
given by
\be (\phi^* \phi)_r = Z_m \phi^* \phi + z_S ~m_H^2 \ee
where $z_S$ is a UV divergent constant.

We note that the renormalized $\A_{r,\mu}$ and $\Phi_r$ are
independent of the constant shift $v_0 = Z_v~v$ of the Higgs field
$\rho$.  However, both in practical loop calculations and in formal
diagrammatic studies, it is extremely convenient to have the tadpole
vanishing.  Therefore, we choose $Z_v$ so that $\vev{\rho} = 0$.

To summarize, the renormalized $\A_{r,\mu}$ and $\Phi_r$ are defined
by
\ba \A_{r,\mu} &\equiv& {1 \over e v^2} {1 \over Z_3} J_\mu = {1 \over
Z_3^{3 \over 2}} {1 \over v^2} B_\mu \left( v_0 + \rho \right)^2 \\
\Phi_r &\equiv& {1 \over v} \left( Z_m \phi^* \phi + z_S~m_H^2 - {v^2
\over 2} \right) \nonumber\\ &=& {1 \over v} \left( Z_m {1 \over 2}
\left( v_0 + \rho \right)^2 + z_S~m_H^2 - {v^2 \over 2} \right) \ea

\subsection{Transverse nature of the Green functions}

We consider the Green functions of $\p_\mu J_\mu$ to understand the
transverse nature of the Green functions.  This is nothing but the
Ward identities except that in renormalizable gauges they are derived
for the Green functions of charged fields.  Here, the Ward identities
are derived for the charge neutral (gauge invariant) fields.  In the
unitary gauge the Feynman rules are simple, and the Ward identities
are most easily derived using Feynman diagrams.  Here only the results
are summarized, leaving a sketch of derivations in Appendix A.  All
the Ward identities are given for the bare fields.

We obtain the following Ward identities:
\ba &&k_\mu \vev{J_\mu (k) J_\nu} = 2 k_\nu \vev{\phi^* \phi}\\
&&k_{1\mu} \vev{J_\mu (k_1) J_\nu (k_2) (\phi^* \phi)} = 2 k_{1\nu}
\vev{(\phi^* \phi) (k_1+k_1) (\phi^* \phi)}\\ &&k_{1\mu} \vev{J_\mu
(k_1) J_\nu (k_2) ~(\phi^* \phi)~(\phi^* \phi)} \nonumber\\ &&\qquad
\qquad= 2 k_{1\nu} \vev{(\phi^* \phi) (k_1+k_2) ~(\phi^* \phi)~(\phi^*
\phi)}\\ &&k_{1\alpha} \vev{J_\alpha (k_1) J_\beta (k_2) J_\gamma
(k_3) J_\delta}\nonumber\\ && \qquad = 2 \Big[ k_{1\beta}
\vev{(\phi^* \phi) (k_1+k_2) J_\gamma (k_3) J_\delta} +
(\rm{permutations}) \Big] \ea

These imply the following structure:
\ba &&\vev{J_\mu (k) J_\nu} = T^{(1)}_{\mu\nu}(k) + 2 \delta_{\mu\nu}
\vev{\phi^* \phi} \label{JJWard}\\ && \vev{J_\mu (k_1) J_\nu (k_2)
~(\phi^*\phi)} = T^{(2)}_{\mu\nu}(k_1,k_2) + 2 \delta_{\mu\nu}
\vev{(\phi^* \phi) (k_1+k_2) (\phi^* \phi)}\\ && \vev{J_\mu (k_1)
J_\nu (k_2) (\phi^* \phi) (p) (\phi^*\phi)} = T^{(3)}_{\mu\nu}
(k_1,k_2,p) \nonumber\\ &&\qquad\qquad\qquad + 2 \delta_{\mu\nu}
\vev{(\phi^* \phi) (k_1+k_2) (\phi^* \phi)(p) (\phi^* \phi)}\\ &&
\vev{J_\alpha (k_1) J_\beta (k_2) J_\gamma (k_3) J_\delta} =
T^{(4)}_{\alpha\beta\gamma\delta}(k_1,k_2,k_3) \\ && ~ + 2 \Big[
\delta_{\alpha\beta} \vev{(\phi^* \phi) (k_1+k_2) J_\gamma J_\delta} +
\delta_{\gamma\delta} \vev{J_\alpha J_\beta (\phi^* \phi) (-k_1 -
k_2)} +~ ... ~\Big] \nonumber\\ && ~ - 4 \Big[ \delta_{\alpha\beta}
\delta_{\gamma\delta} \vev{(\phi^* \phi) (k_1+k_2) (\phi^* \phi)} +
~...~ \Big] \label{four}\ea
where $T^{(i)}$ tensors are all transverse:
\ba k_\mu T^{(1)}_\mu (k) &=& 0 \\
k_{1,\mu} T^{(2)}_{\mu\nu}(k_1,k_2) &=& 0, \quad
k_{2,\nu} T^{(2)}_{\mu\nu}(k_1,k_2) = 0 \\
k_{1,\mu} T^{(3)}_{\mu\nu}(k_1,k_2,p) &=& 0, \quad
k_{2,\nu} T^{(3)}_{\mu\nu}(k_1,k_2,p) = 0 \\
k_{1,\alpha} T^{(4)}_{\alpha\beta\gamma\delta} (k_1,k_2,k_3) &=& 0,
~~... \ea

The above results can be easily generalized to an arbitrary Green
function of $J$'s and $(\phi^* \phi)$'s.  In general we find
\be \vev{J_{\mu} ~...~ (\phi^*\phi)~...~} = ({\rm transverse~part}) +
~...~ \label{general}\ee
where the terms denoted by dots are Green functions in which pairs of
$J$'s are replaced by $(\phi^*\phi)$'s.

\subsection{Photon propagator}

Eq.~(\ref{JJWard}) gives the longitudinal part of $\vev{J_\mu J_\nu}$
explicitly, but we can analyze its structure in much more detail.
Especially we can express $\vev{J_\mu J_\nu}$ in terms of the full
propagator $\vev{B_\mu B_\nu}$.  (See Appendix B for a sketch of the
derivation.)  Let us first denote the self-energy of $B_\mu$ by
\be \Sigma_{\mu\nu} (k) = \delta_{\mu\nu} \Sigma_1 (k^2) + {k_\mu
k_\nu \over m_0^2} \Sigma_2 (k^2) \ee
where $m_0^2 \equiv e_0^2 v_0^2$, so that the full propagator is given
by
\be \vev{B_\mu (k) B_\nu} = {1 \over k^2 + \Sigma_1} \left(
\delta_{\mu\nu} - {k_\mu k_\nu \over k^2} \right) + {k_\mu k_\nu \over
k^2} {m_0^2 \over m_0^2 \Sigma_1 + k^2 \Sigma_2} \label{fullBB}\ee

Now we can show diagrammatically the following relation:
\be \vev{J_\mu J_\nu} = {1 \over e_0^2} \Sigma_{\mu\alpha}
\vev{B_\alpha B_\beta} \Sigma_{\beta\nu} - {1 \over e_0^2}
\Sigma_{\mu\nu} + 2 \delta_{\mu\nu} \vev{\phi^* \phi} \label{JJbyBB}
\ee
The two-point function of the current is then found to have the
following structure:
\be \vev{J_\mu (k) J_\nu} = {1 \over e_0^2} \Bigg[ {(k^2)^2 \over
k^2 + \Sigma_1} \left( \delta_{\mu\nu} - {k_\mu k_\nu \over
k^2}\right) - k^2 \delta_{\mu\nu} + k_\mu k_\nu \Bigg]
+ 2 \delta_{\mu\nu} \vev{\phi^* \phi} \ee
Note that this depends only on the transverse part $\Sigma_1$ but not
on the longitudinal part $\Sigma_2$.

Since ${1 \over \sqrt{Z_3}} F_{\mu\nu}$ is renormalized, the
transverse part of $\vev{B_\mu B_\nu}$ can be renormalized by the
factor ${1 \over Z_3}$.  Hence,
\be {1 \over Z_3} {1 \over k^2 + \Sigma_1} \ee
is free of UV divergences.  Therefore, for the renormalized current
$J_{r,\mu}$, we obtain
\ba &&\vev{J_{r,\mu} (k) J_{r,\nu}} = {1 \over e^2} {1 \over Z_3 (k^2
 + \Sigma_1)} (k^2)^2 \left( \delta_{\mu\nu} - {k_\mu k_\nu \over
 k^2}\right) \nonumber\\ && - {1 \over e^2} {1 \over Z_3} \left( k^2
 \delta_{\mu\nu} - k_\mu k_\nu \right) + 2 \delta_{\mu\nu} \left[ {1
 \over Z_3 Z_m} \vev{(\phi^* \phi)_r} - {z_S \over Z_3 Z_m} m_H^2
 \right] \label{JJ}\ea

\subsection{Renormalization of the Green functions}

In coordinate space the Green functions of the renormalized
$J_{r,\mu}$'s and $(\phi^*\phi)_r$'s are UV finite as long as we keep
all the space points distinct.  When two or more points coincide, a
non-integrable singularity can be generated, and this gives rise to a
UV singularity in the Fourier transform.

The renormalized current $J_{r,\mu}$ has scale dimension 3, and we
expect its two-point function to have a singularity which is a
polynomial in momentum up to second order.  The formula (\ref{JJ}) we
derived above gives a precise structure of the UV divergence.  Hence,
a renormalized two-point function is defined by
\ba \vev{J_{r,\mu} (k) J_{r,\nu}}_{ren} &\equiv& \vev{J_{r,\mu} (k)
J_{r,\nu}} + {1 \over e^2} \left( {1 \over Z_3} - 1 \right) (k^2
\delta_{\mu\nu} - k_\mu k_\nu) \nonumber\\ && - 2 \delta_{\mu\nu}
\left[ \left({1 \over Z_3 Z_m} - 1 \right) \vev{(\phi^*\phi)_r} -
\left({z \over Z_3 Z_m}\right) m_H^2 \right]\nonumber\\ &=& {1 \over
e^2} \left({1 \over Z_3 (k^2 + \Sigma_1)} (k^2)^2 - k^2 \right)
\left(\delta_{\mu\nu} - {k_\mu k_\nu \over k^2}\right) \nonumber\\ &&
\qquad\qquad + 2 \delta_{\mu\nu} \vev{(\phi^*\phi)_r} \ea

The scale dimension of $(\phi^* \phi)_r$ is 2, and its two-point
function has a constant singularity independent of momentum.  Thus, we
can define
\be \vev{(\phi^*\phi)_r (p) (\phi^*\phi)_r}_{ren} \equiv
\vev{(\phi^*\phi)_r (p) (\phi^*\phi)_r} - {\cal D} \ee
where ${\cal D}$ is a UV divergent constant.\footnote{Actually one can
show ${\cal D} = - 2 z_S$ to all orders in perturbation theory.}

Let us now consider the three-point function $\vev{J_{r,\mu} J_{r,\nu}
(\phi^*\phi)_r}$.  In coordinate space the product of two currents
gives a dimension four singularity proportional to the scalar
$(\phi^*\phi)_r$, and this gives rise to the only non-integrable
singularity.  Hence, the transverse part is multiplicatively
renormalizable, and we obtain
\ba &&\vev{J_{r,\mu}(k_1)~J_{r,\nu}(k_2)~(\phi^*\phi)_r}_{ren} \equiv
\vev{J_{r,\mu}~J_{r,\nu}~(\phi^*\phi)_r} \nonumber\\ && ~- 2
\delta_{\mu\nu} \left({1 \over Z_3^2 Z_m} - 1 \right)
\vev{(\phi^*\phi)_r (k_1+k_2) (\phi^*\phi)_r} - 2 \delta_{\mu\nu}
{\cal D}\nonumber\\ &&= {Z_m \over Z_3^2} T^{(2)}_{\mu\nu} + 2
\delta_{\mu\nu} \vev{(\phi^*\phi)_r(k_1+k_2)~(\phi^*\phi)_r}_{ren}\ea

The product of two $(\phi^*\phi)$'s contains a constant singularity
${\cal D}$ as we have seen above, but this singularity does not affect
the three-point function of $(\phi^*\phi)$ for generic external
momenta.  Hence, we do not need any subtraction, and we obtain
\be \vev{(\phi^*\phi)_r~(\phi^*\phi)_r~(\phi^*\phi)_r}_{ren} =
\vev{(\phi^*\phi)_r~(\phi^*\phi)_r~(\phi^*\phi)_r} \ee
Similarly, the four-point (or higher-point) function of
$(\phi^*\phi)$'s does not require further subtractions:
\be
\vev{(\phi^*\phi)_r~(\phi^*\phi)_r~(\phi^*\phi)_r~(\phi^*\phi)_r}_{ren}
= \vev{(\phi^*\phi)_r~(\phi^*\phi)_r~(\phi^*\phi)_r~(\phi^*\phi)_r}
\ee

In the Green function of two $J$'s and two $(\phi^*\phi)$'s, the only
singularity comes from two $J$'s coincident in space, producing a
scalar $(\phi^*\phi)$.  Hence, the transverse part is multiplicatively
renormalizable, and we obtain
\ba
&&\vev{J_{r,\mu}(k_1)~J_{r,\nu}(k_2)~(\phi^*\phi)_r~(\phi^*\phi)_r}_{ren}
\equiv
\vev{J_{r,\mu}~J_{r,\nu}~(\phi^*\phi)_r~(\phi^*\phi)_r}\nonumber\\ &&
\qquad - 2 \delta_{\mu\nu} \left( {1 \over Z_3^2 Z_m} - 1 \right)
\vev{(\phi^*\phi)_r~(\phi^*\phi)_r~(\phi^*\phi)_r} \nonumber\\ &&=
{Z_m^2 \over Z_3^2} T^{(3)}_{\mu\nu} + 2 \delta_{\mu\nu}
\vev{(\phi^*\phi)_r (k_1+k_2) ~(\phi^*\phi)_r~(\phi^*\phi)_r}_{ren}
\ea

Finally, we consider the four-point function of $J$'s.  This has two
causes of UV singularities: one coming from a pair of coincident $J$'s
producing a $\phi^*\phi$, and another coming from four coincident
$J$'s producing an identity operator.  On the other hand
Eq.~(\ref{four}) gives
\ba
&&\vev{J_{r,\alpha}(k_1)~J_{r,\beta}(k_2)~J_{r,\gamma}(k_3)~J_{r,\delta}}
= {1 \over Z_3^4} T^{(4)}_{\alpha\beta\gamma\delta} \nonumber\\ &&
\quad +{2 \over Z_3^2 Z_m} \left( \delta_{\alpha\beta}
\vev{(\phi^*\phi)_r (k_1+k_2) J_{r,\gamma} J_{r,\delta}} +
~...~\right)\nonumber\\ && \quad - {4 \over Z_3^4 Z_m^2} \left(
\vev{(\phi^*\phi)_r (k_1+k_2) (\phi^*\phi)_r} + ~...~\right) \ea
Therefore, the transverse part is multiplicatively renormalizable, and
we can renormalize the four-point function as
\ba
&&\vev{J_{r,\alpha}(k_1)~J_{r,\beta}(k_2)~J_{r,\gamma}(k_3)~J_{r,\delta}}_{ren}
\equiv \vev{J_{r,\alpha}~J_{r,\beta}~J_{r,\gamma}~J_{r,\delta}}
\nonumber\\ && \quad - 2 \left( {1 \over Z_3^2 Z_m}-1 \right) \left (
\delta_{\alpha\beta} \vev{(\phi^*\phi)_r~J_{r,\gamma}~J_{r,\delta}} +
~...~ \right) \nonumber\\ &&\quad + 4 \left\lbrace \left({1 \over
Z_3^4 Z_m^2} - 1 \right) - 2 \left({1 \over Z_3^2 Z_m} - 1 \right)
\right\rbrace \left( \delta_{\alpha\beta} \delta_{\gamma\delta}
\vev{(\phi^* \phi)_r~(\phi^*\phi)_r} + ~...~\right) \nonumber\\
&&\quad - 4 {\cal D} \left( \delta_{\alpha\beta} \delta_{\gamma\delta}
+ ~...~\right) \nonumber\\ &&= {1 \over Z_3^4}
T^{(4)}_{\alpha\beta\gamma\delta} + 2 \left ( \delta_{\alpha\beta}
\vev{(\phi^*\phi)_r~J_{r,\gamma}~J_{r,\delta}}_{ren} + ~...~\right)
\nonumber\\ &&\qquad - 4 \left(
\delta_{\alpha\beta}\delta_{\gamma\delta} \vev{(\phi^*\phi)_r
(k_1+k_2)~(\phi^*\phi)_r}_{ren} +~...~\right) \ea

The above results generalize easily for an arbitrary renormalized
Green function of $J$'s and $(\phi^*\phi)$'s.  From
Eq.~(\ref{general}), we find
\be \vev{J_{r,\mu} ~...~(\phi^*\phi)_r ~...~}_{ren} =
(\rm{renormalized~transverse~part}) + ~...~ \ee
where the transverse part is multiplicatively renormalized, and the
dotted part is a linear combination of lower point renormalized Green
functions.

\subsection{1-loop results}

We have explicitly computed the UV divergent part of all the two-point
and three-point functions and the four-point function $\vev{\A \A \A
\A}$, and we have verified the structure of UV divergences found in
the previous subsection.  Besides the renormalization constants $Z_3$,
$Z_m$, $Z_\lambda, z$ given in Eqs.~(\ref{z3},\ref{zm},\ref{zlambda}),
we have obtained\footnote{Note that the relation ${\cal D} = - 2 z_S$
alluded to in the previous footnote is satisfied at 1-loop.}
\be z_S = - {1 \over 2} \ln \Lambda, \quad {\cal D} = \ln \Lambda \ee
No other renormalization constants are necessary to renormalize higher
point Green functions.

\section{S-matrix elements}

Before concluding this paper, let us briefly discuss the implications
of our results to the calculation of the S-matrix elements in the
unitary gauge.

The first step is to calculate the off-shell Green function
\be \vev{\A_{r,\mu} ~...~ \Phi_r ~...~} \ee
using $\A_{r,\mu} = {1 \over e v^2} J_{r,\mu}$ and $\Phi_r = {1 \over
v} \left( (\phi^*\phi)_r - {v^2 \over 2} \right)$ as interpolating
fields.  This is given as the sum of a UV finite transverse part and a
linear combination of lower point Green functions with UV divergent
coefficients.  The UV divergent part is irrelevant to the S-matrix,
since it does not have the desired pole structure as we take the
external momenta to the mass shell.  Hence, only the transverse part
contributes to the S-matrix element, and it is UV finite.

In practice we do not need to use $\A_r$ and $\Phi_r$ as interpolating
fields; according to the general theory of S-matrix, we should be able
to use any interpolating fields.  The simplest choice of the
interpolating fields is definitely $B_\mu$ for the photon and $\rho$
for the Higgs.  Their Green functions are not multiplicatively
renormalizable, but by normalizing $B_\mu$ and $\rho$ properly we can
obtain the same S-matrix elements as from the Green functions of
$\A_r$'s and $\Phi_r$'s.

The propagator of $B_\mu$ is given by Eq.~(\ref{fullBB}), and only its
transverse part is renormalizable:
\be {1 \over Z_3} \vev{B_\mu (k) B_\nu} = {1 \over Z_3 (k^2 +
\Sigma_1)} \left( \delta_{\mu\nu} - {k_\mu k_\nu \over k^2} \right) +
k_\mu k_\nu (\rm{UV~divergent}) \ee
In perturbation theory the longitudinal part has no single particle
pole, and we can use ${1 \over \sqrt{Z_3}}B_\mu$ as an interpolating
field of the massive photon.  As for $\rho$, its two-point function is
not renormalizable, and the residue $Z_\rho$ at the single particle
pole is UV divergent:
\be \vev{\rho (p) \rho} \to Z_\rho {1 \over p^2 + m_H^2} \quad {\rm as}~p^2
\to - m_H^2 \ee
Hence, we must use ${\rho \over \sqrt{Z_\rho}}$ as the interpolating
field of the Higgs.

Therefore, for the Green function with $n_B$ $B$'s and $n_\rho$
$\rho$'s,
\be {1 \over Z_3^{n_B \over 2} Z_\rho^{n_\rho\over 2}} \vev{B_\mu
~...~\rho~...~} \ee
has the same UV finite pole part as
$\vev{\A_{r,\mu}~....~\Phi_r~...~}$ as we take the mass shell limit of
the external momenta.\footnote{The two pole parts can differ by a
finite multiplicative constant since the single particle residue of
$\vev{\Phi_r \Phi_r}$ is UV finite but not necessarily $1$.}  The
possibility of calculating the S-matrix elements from the Green
functions of $B$'s and $\rho$'s is thus assured.

\section{Concluding remarks}

We hope to have convinced the reader of the off-shell
renormalizability of the unitary gauge for the abelian Higgs model.
In perturbative calculations in the unitary gauge, we certainly
encounter unfamiliar UV divergences, but our analysis in this paper
clarifies the origin of these divergences: mixing and high scale
dimensions of the fields.  We have also seen how these divergences
disappear in the physical limit.  Now that we understand the nature
and the reason of these divergences, the unitary gauge should be
promoted to the same rank as the covariant and $R_\xi$ gauges.

The extension of the present work to non-abelian Higgs theories should
be straightforward.  Especially for those non-abelian Higgs theories
with no massless gauge boson (for example, the SU(2) theory with a
Higgs doublet), we expect no qualitative change in the analysis.  For
those theories with massless gauge bosons, however, a gauge fixing is
necessary, and we will need to modify the analysis to accommodate the
gauge fixing.

\vspace{0.5cm}
\noindent{\Large \bf Acknowledgment}

\vspace{0.2cm}
This paper was written during my visit to the Department of Physics of
Penn State University.  I would like to thank Prof.~Murat G\"unayden
for hospitality.  This work was supported in part by the Grant-In-Aid
for Scientific Research from the Ministry of Education, Science, and
Culture, Japan (\#11640279).

\appendix

\section{Ward identities}

We sketch a diagrammatic derivation of the Ward identities given in
the main text.  We consider bare quantities, and $v_0$ is determined
so that $\vev{\rho} = 0$, and $m_0^2 \equiv e_0^2 v_0^2$.  The choice
of the vanishing tadpole is not essential, but it simplifies the
diagrammatic analysis considerably.  The propagator of the photon is
given by
\be D_{\mu\nu} (k) \equiv \vev{B_\mu (k) B_\nu}_{tree} =
{\delta_{\mu\nu} + {k_\mu k_\nu \over m_0^2} \over k^2 + m_0^2} \ee
This satisfies
\be k_\mu D_{\mu\nu} (k) = {k_\nu \over m_0^2} \label{dD}\ee
and we write this relation diagrammatically as
\begin{center}
\epsfig{file=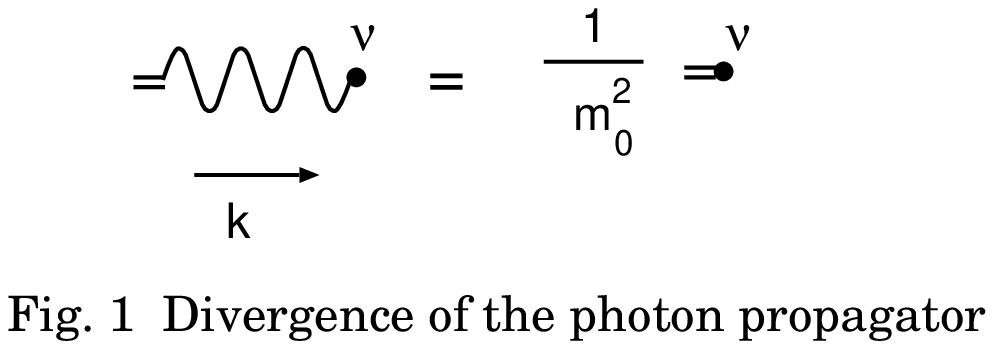, height=3cm}
\end{center}
where a double line represents a derivative factor $k_\mu$, and a dot
represents a Kronecker delta $\delta_{\mu\nu}$.  The Feynman rules of
the vertices involving photons are given by
\begin{center}
\epsfig{file=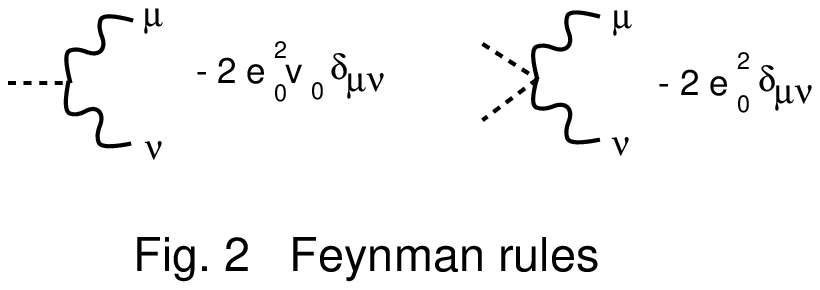, height=3cm}
\end{center}
where a broken line represents a Higgs propagator.

We first consider the two-point function of $\A_\mu$.  We obtain the
diagrammatic expression given below, in which a cross denotes
amputation of the external propagator:
\begin{center}
\epsfig{file=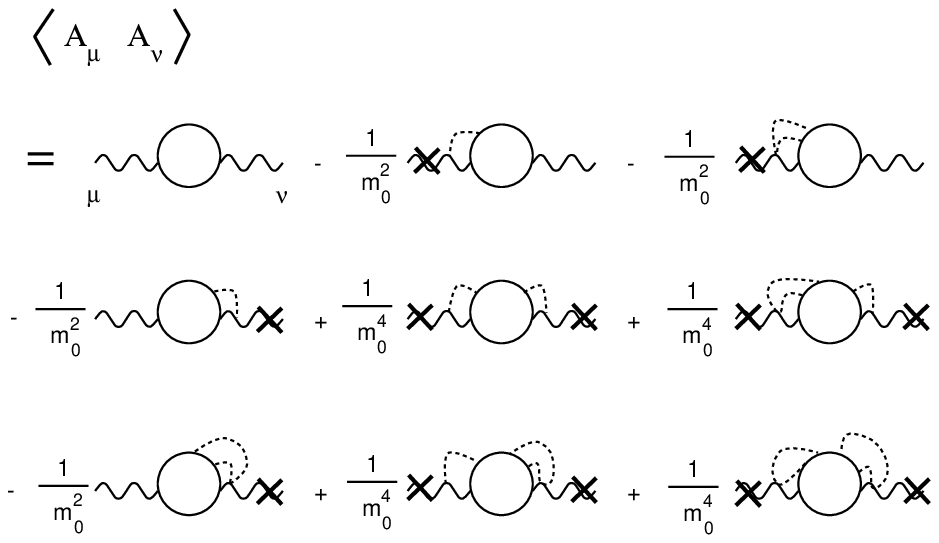, height=7cm}
\end{center}
Taking into account the vanishing tadpole, we now make the following
observation: 
\begin{center}
\epsfig{file=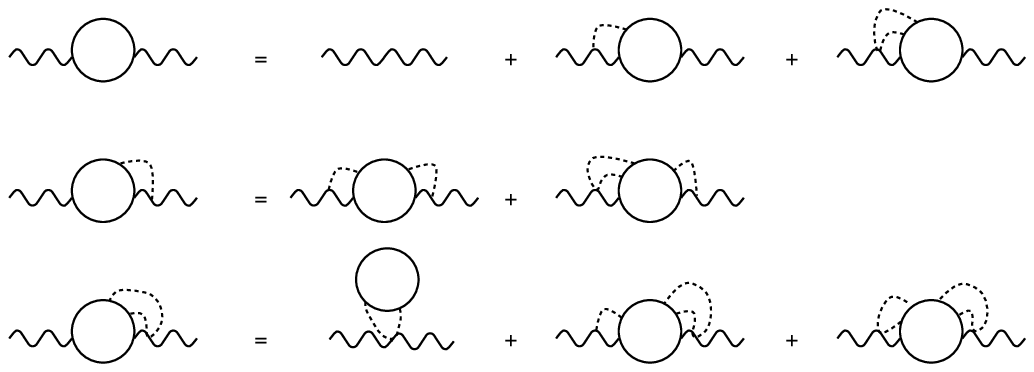, height=4.5cm}
\end{center}

Now, using Eq.~(\ref{dD}), we find a lot of cancellations, and we
eventually obtain
\begin{center}
\epsfig{file=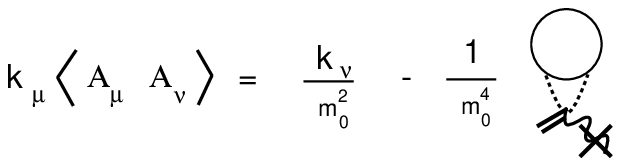, height=2cm}
\end{center}
which implies
\be k_\mu \vev{\A_\mu (k) \A_\nu} = {2 k_\nu \over m_0^2 v_0^2}
\vev{\phi^* \phi} \ee

For the Green function $\vev{\A_\mu (k) \A_\nu (k') ~\Phi ~...~\Phi}$
with an arbitrary number of $\Phi$'s, the derivation of the Ward
identity goes the same way as above.  The extra insertions of $\Phi$'s
does not change the nature of derivation.  We only need to pay
attention to the two $\A$'s:
\begin{center}
\epsfig{file=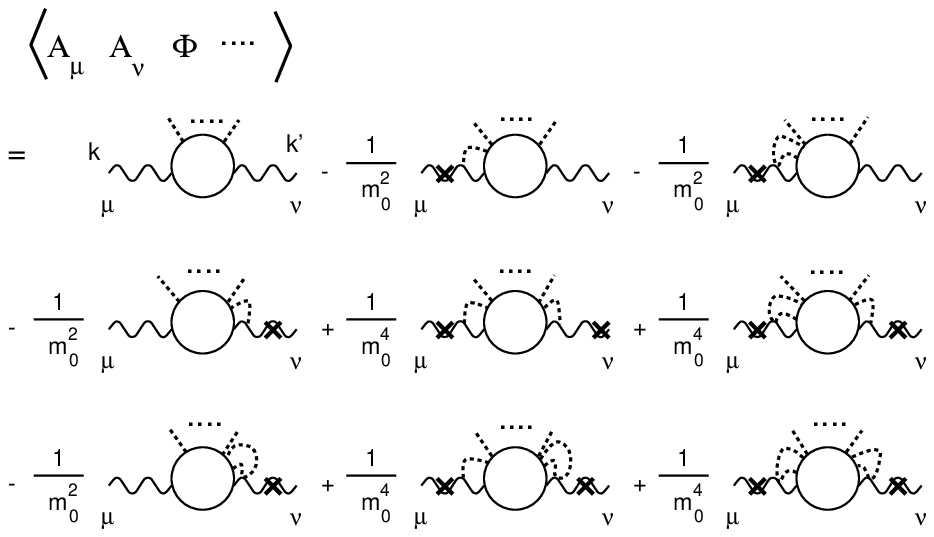, height=6cm}
\end{center}
Using Eq.~(\ref{dD}) and diagrammatic identities similar to those for
the propagator, we obtain
\begin{center}
\epsfig{file=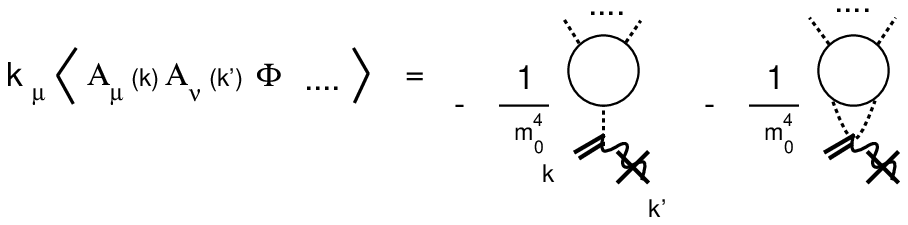, height=3cm}
\end{center}
which implies
\be k_\mu \vev{\A_\mu (k) \A_\nu (k') ~\Phi ~...~} = {2 \over m_0^2
v_0} k_\nu \vev{\Phi (k+k') \Phi~...~} \ee

With more insertions of $\A$'s, the derivation goes similarly.  In
general we obtain
\ba &&k_{1\mu_1} \vev{\A_{\mu_1} (k_1) \A_{\mu_2} (k_2)
~...~\A_{\mu_M} (k_M) ~\Phi ~...~}\nonumber\\ &=& {2 \over m_0^2 v_0}
\sum_{m=2}^M k_{1\mu_m} \vev{ \A_{\mu_2} (k_2) ~...~\Phi
(k_1+k_m)~...~\A_{\mu_M} (k_M) ~\Phi~...~} \ea
In the main text, the Ward identities are rewritten for $J$'s and
$(\phi^*\phi)$'s instead of $\A$'s and $\Phi$'s.

\section{The relation between $\vev{B_\mu B_\nu}$ and $\vev{\A_\mu
\A_\nu}$}

With a more detailed diagrammatic analysis, we can derive an explicit
relation between $\vev{B_\mu~B_\nu}$ and $\vev{\A_\mu~\A_\nu}$.  Let
us denote the self-energy correction to the propagator by
$\Pi_{\mu\nu}$ so that the full propagator $\vev{B_\mu~B_\nu}$ can be
written recursively as
\be \vev{B_\mu~B_\nu} = D_{\mu\nu} - D_{\mu \alpha} \Pi_{\alpha\beta}
\vev{B_\beta~B_\nu} \ee
The self-energy correction can be written in terms of 1PI diagrams as
follows:
\begin{center}
\epsfig{file=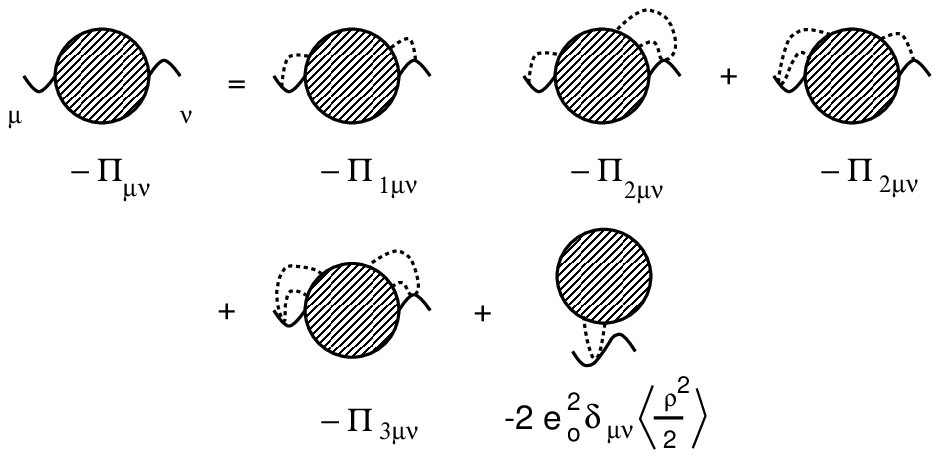, height=4cm}
\end{center}

We now observe the following diagrammatic identities:
\begin{center}
\epsfig{file=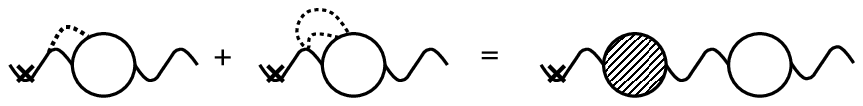, height=1cm}
\end{center}
\begin{center}
\epsfig{file=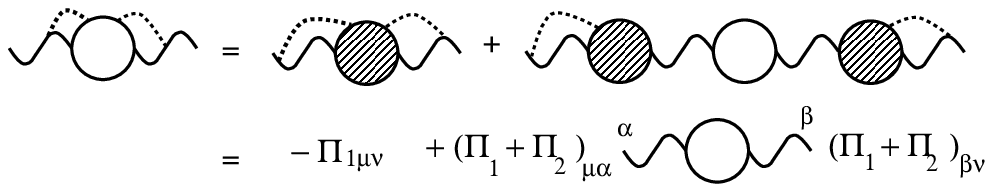, height=2cm}
\end{center}
\begin{center}
\epsfig{file=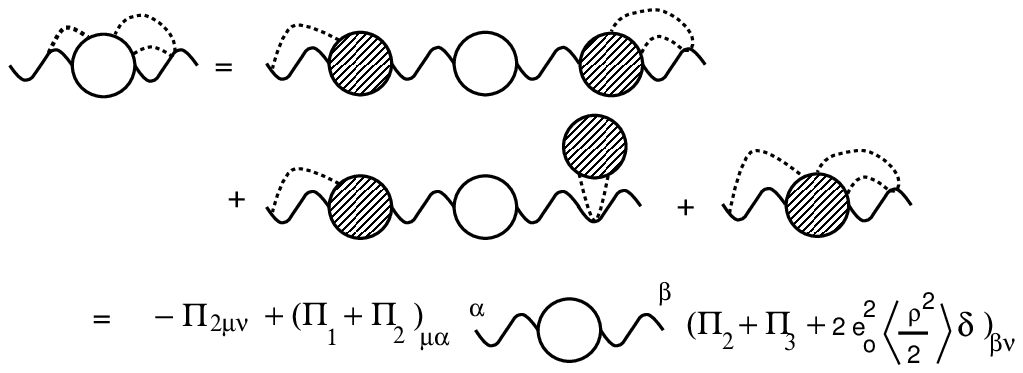, height=4cm}
\end{center}
\begin{center}
\epsfig{file=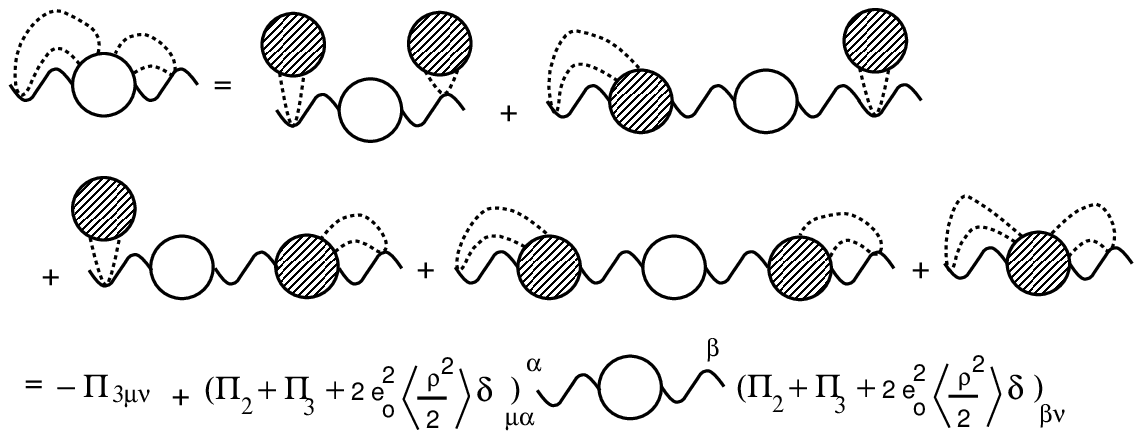, height=4cm}
\end{center}
Using these and the diagrammatic expression of $\vev{\A\A}$ given in
Appendix A, we obtain
\be \vev{\A_\mu~\A_\nu} = {1 \over m_0^4} \left[ \Sigma_{\mu\alpha}
\vev{B_\alpha~B_\beta} \Sigma_{\beta\nu} - \Sigma_{\mu\nu} + 2 e_0^2
\delta_{\mu\nu} \vev{\phi^*\phi} \right] \ee
where
\be \Sigma_{\mu\nu} \equiv m_0^2 \delta_{\mu\nu} + \Pi_{\mu\nu} \ee
Replacing $e_0 v_0^2 \A$ by $J$, we obtain the relation (\ref{JJbyBB})
in the main text.

\end{document}